\documentstyle[prl,aps,epsf,multicol]{revtex}

\begin{document}
\draft
\tighten

\title{A Discotic Disguised as a Smectic: A Hybrid Columnar Bragg Glass}

\author{Karl Saunders$^1$, Leo Radzihovsky$^2$, 
John Toner$^1$}
\address{$^1$ Dept. of Physics,
Materials Science Inst., and Inst. of Theoretical Science, University
of Oregon, Eugene, OR 97403}
\address{$^2$ Department of Physics, University of Colorado,
Boulder, CO 80309}

\date{\today}
\maketitle
\begin{abstract}
  
  We show that discotics, lying deep in the columnar phase, can
  exhibit an x-ray scattering pattern which mimics that of a somewhat 
  unusual smectic liquid crystal. This exotic, new glassy phase of columnar 
  liquid crystals, which we call a ``hybrid columnar Bragg glass'', can be
  achieved by confining a columnar liquid crystal in an anisotropic
  random environment of e.g., strained aerogel. Long-ranged
  orientational order in this phase makes {\em single domain} x-ray
  scattering possible, from which a wealth of information could be
  extracted. We give detailed quantitative predictions for the
  scattering pattern in addition to exponents characterizing anomalous
  elasticity of the system.

\end{abstract}

%\twocolumn
\vspace{-0.5cm}
\begin{multicols}{2}
\narrowtext

\begin{figure}[bth] 
\centering
\setlength{\unitlength}{1mm} 
\begin{picture}(15,42)(0,-12)
\put(-35,-45){\begin{picture}(20,20)(0,0) 
\includegraphics{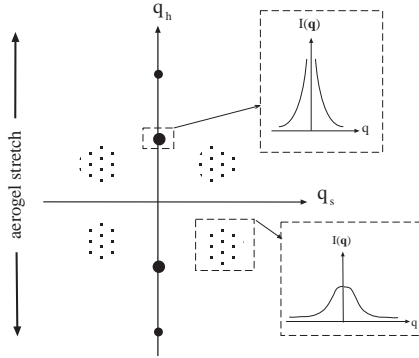} 
\end{picture}} 
\end{picture} 
\caption{X-ray scattering pattern in the $\perp$ plane for a class of hybrid
columnar Bragg glass.} 
\label{scattering} 
\end{figure}

Until now, the x-ray scattering pattern given in Fig.\ref{scattering}
would be identified with a system in a somewhat unusual smectic phase
with short-ranged translational order and long-ranged orientational
order {\em within} the smectic layers; i.e., a smectic composed of
{\it nematic}, rather than {\it liquid}, layers.  The set of on-axis
quasi-sharp Bragg peaks along $q_h$ is a signature of the
quasi-long-ranged translational order (i.e., the periodicity of the
layering) that is characteristic of the bulk smectic phase. The
presence of the other, broadened, peaks and the azimuthal anisotropy
about the $q_h$ axis respectively indicate the incipient {\it
short-ranged} translational order and the long-ranged orientational
order {\it within} the smectic layers oriented perpendicular to $q_h$.

In this Letter we predict the existence of a remarkable new ``hybrid columnar
Bragg\cite{NS,RFXY} glass'' phase (HCBG) which, despite differing
fundamentally from the smectic phase described above, shares the same
qualitative scattering pattern illustrated in Fig.\ref{scattering}. Such 
mimickry of one phase by a completely different
other phase is unprecedented.

\begin{figure}[bth]
\centering
\setlength{\unitlength}{1mm}
\begin{picture}(30,42)(0,0)
\put(-30,-33){\begin{picture}(20,20)(0,0)
\includegraphics{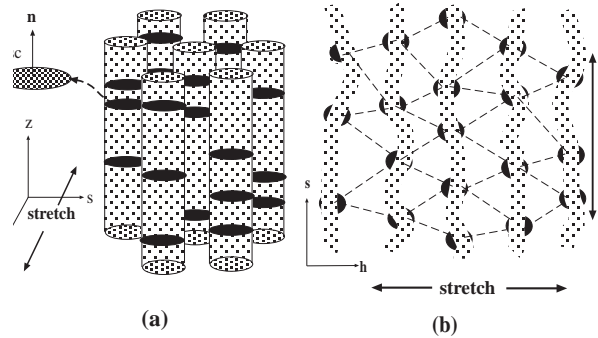}
\end{picture}}
\end{picture}
\caption{(a) Alignment of the lattice for a uniaxial stretch. 
(b) Schematic of the distorted lattice in real space.}
\label{discoticside}
\end{figure}

Columnar phases in pure, bulk (i.e., quenched-disorder free) liquid
crystals are phases that have long-ranged translational order in
two-directions, and short-ranged translational order (i.e.,
liquid-like correlations) in the third. I.e., they are
regular two-dimensional lattices of one-dimensional 
liquid columns (Fig.\ref{discoticside}(a)). In this Letter,
we show that when such a system is confined in an {\em anisotropic}
quenched random environment, e.g., strained aerogel\cite{Clarke}, it becomes
translationally disordered, but remains topologically ordered (i.e.,
free of topological defects like dislocations). This novel state is
the HCBG.

Like the smectic phase, the HCBG has translational order that is
quasi-long-ranged in one direction and short-ranged in another, as implied by
Fig.\ref{scattering}. As illustrated in Fig.\ref{discoticside}(b), the columns
remain in roughly equidistant rows perpendicular to the stretching direction,
but lose long-ranged translational order within each row. Nevertheless, the
hexagonal orientational order is preserved, albeit uniaxially distorted due to
the strectch.  However, there are a number of fundamental differences between
the two phases.  First, unlike the smectic phase, the exponent $\eta_{\bf G}$
characterizing the shape of the quasi-long-ranged translational order peaks at
${\bf G}$, which are given by $I({\bf q})\propto |{\bf q}-{\bf G}|^{ -3+
{\eta_{G} } }$, 
% 
%\begin{eqnarray} 
%I({\bf q})\propto |{\bf q}-{\bf G}|^{-3+\eta({\bf G})} 
%\label{I} 
%\end{eqnarray} 
% 
is independent of temperature.  This property would allow the two phases'
scattering patterns to be distinguished through comparison of the lineshapes
as temperature is varied. Secondly, the correlations of the quasi-long-ranged
order in the HCBG scale isotropically in space, in contrast to the well-known
strongly anisotropic scaling of these correlations in the smectic.  The third
and most crucial difference between the two phases is their topolological
order, which distinguishes their elasticities but not their scattering. 
Specifically, the absence of the extra direction of long-ranged
translational order, in the smectic phase, is caused by free topological
defects, namely unbound dislocations with Burgers vectors along the smectic
planes. Although the HCBG also exhibits translational order that is 
quasi-long-ranged in one
direction and short-ranged in another, at long length scales it is
distinguished from the smectic by being free of these unbound dislocations,
which would otherwise destroy the columnar phase topology of HCBG. One
important experimental consequence of this absence of free dislocations is
that the HCBG retains elastic resistance to distortions in the extra
direction, albeit, as we discuss below, of a very strange, anomalous sort.

Of course, for {\em sufficiently} strong disorder 
free dislocations {\em will} eventually proliferate. 
The anisotropy, imposed by the strained aerogel, leads to
the interesting possibility that dislocations with their Burgers
vectors in the soft direction may unbind (with increasing disorder) 
before those with their Burgers vectors in the hard direction, leading 
to the sequence of disorder-driven phase transitions : HCBG $\rightarrow$ m=1 smectic Bragg
glass{\cite {JSRT}} $\rightarrow$ nematic elastic glass\cite{RTpr} with
increasing aerogel density.

The rest of this Letter gives a more detailed theoretical description of
the HCBG phase, including x-ray correlation lengths and universal
exponents characterizing the anomalous elasticity. We relegate
the technical details to a future publication\cite{SJRT}.  

Our model for a columnar phase consists of
disc-shaped molecules with normals aligned along the $\hat {\bf z}$
direction. The discs form a hexagonal lattice in the $xy$ ($\perp$)
plane and have liquid like correlations along $\hat {\bf z}$ as
illustrated in Fig.\ref{discoticside}(a).  We assume, and verify a
posteriori, that despite considerable distortion, for sufficiently
weak quenched disorder, the columnar phase topology is stable, i.e.
our discotic liquid crystal remains free of unbound dislocations.
Consequently this system can be described within an elastic theory,
with a two-component ($x$ and $y$) lattice site displacement vector
$\bf u(\bf r)$ and the discotic director $\hat{\bf n}(\bf r)$ (the normal to
the discs) as the only important long length-scale degrees of freedom. The
disordering tendency of the aerogel is two-fold: the strands act both
to randomly pin the columnar lattice ($\bf u(\bf r)$) and to distort
the orientations of the disc normals ($\hat{\bf n}(\bf r)$).  Our starting
Hamiltonian is that of a pure hexagonal discotic in {\em isotropic}
aerogel
\begin{eqnarray}
H&=&\int_{\bf r} \bigg[ {B_\perp\over2} |\partial_z {\bf u} - {\bf
{\delta}} {\bf n}|^2 +{1\over 2}\lambda u_{ii}^\perp u_{jj}^\perp +\mu
u_{ij}^\perp u_{ij}^\perp\nonumber\\ &+&{\cal R}e \sum_i V_i({\bf
r})e^{i {\bf G}_i \cdot {\bf u}({\bf r})} -({\bf g}({\bf
r})\cdot{\hat{\bf n}})^2 \bigg] + H_F [\hat{\bf n}]\;.
\label{H}
\end{eqnarray}
where $u_{ij}^\perp = {1\over2}(\partial_i^\perp u_j +\partial_j^\perp
u_i - \partial_k u_i \partial_k u_j)$ is the rotationally invariant
symmetric strain tensor, $ \delta {\bf n}({\bf r}) \equiv {\bf \hat
  n}({\bf r}) - \hat {\bf z}$, the $B_\perp$ term reflects the
tendency of the molecular director (disc normal) $\hat{\bf n}(\bf r)$ to lie
along the local tangent ${\bf \hat t} \approx {\bf \hat z} +
\partial_z {\bf u} $ to the liquid-like columns, $H_F [{\hat{\bf n}}]$
is the Frank free energy of the molecular directors, and $V_i({\bf
  r})$ is a complex random pinning potential that couples to lattice
site fluctuations along the reciprocal lattice basis vector, ${\bf
  G}_i$. At long length-scales its correlations can be accurately
represented as zero-mean with {\em short-ranged}, Gaussian statistics:
$\overline{V_i({\bf r})V_j^*({\bf r'})} = \tilde{\Delta}_V
\delta_{ij}^{\perp} \delta^d({\bf r}-{\bf r'})$,\cite{RTpr} 
throughout this paper $\overline{x}$ denotes a quenched average
over the disorder of the quantity $x$, while $\langle x\rangle$
denotes a thermal average. The last term
describes the tendency of the disc normals $\hat{\bf n}({\bf r})$ to
align along the random local aerogel strand directed along ${\bf g} ({\bf r})$. 
This ``random tilt'' disorder is described by {\em short-ranged,
  isotropic} correlations $\overline{g_i({\bf r})g_j({\bf r'})} =
1/2\sqrt{\Delta}\delta_{ij}\delta^d({\bf r}-{\bf r'})$\cite{RTpr}. 
$\Delta_V$ and $\Delta$ are phenomenological parameters
which, in the simplest microscopic model, are proportional
to the aerogel density only, $\rho_A$. As for smectics
\cite{RTpr}, these two types of
disorder have important long distance effects.
A detailed analysis\cite{SJRT,RTpr} has shown that this system
exhibits a columnar Bragg glass phase with only short-ranged translational order. 
However, this changes if the aerogel
is anisotropic. Aerogel anisotropy could be realized e.g., by
applying a strain to the strands. For heterotropic alignment between
the disc normals and strands (assumed throughout) a uniaxial {\em
compression} will lead to a phase in the same universality class as
the Bragg glass phase of an Abrikosov flux lattice\cite{NS,RFXY}, with
quasi-long-ranged translational order in both directions of the
$\perp$ plane\cite{SJRT}. The more interesting HCBG
phase with quasi-long-ranged order in only one $\perp$ direction can
be obtained by applying a uniaxial {\em stretch} to the strands. For
homeotropic alignment of strands and disc normals the two phases
reverse with respect to stretch and compression leaving all of our
other predictions unchanged.
%
%\begin{figure}[bth]
%\centering
%\setlength{\unitlength}{1mm}
%\begin{picture}(36,85)(0,0)
%\put(-15,-70){\begin{picture}(80,80)(0,0)
%\special{psfile=discoticside.ps vscale=100 hscale=100}
%\end{picture}}
%\end{picture}
%\caption{Model for the hexagonal discotic.}
%\label{fig2}
%\end{figure}
%

Uniaxial stretch of the aerogel strands(along $\hat {\bf e_h}$ causes
the disc normals to align $\perp$ to the axis of stretch
(Fig.\ref{discoticside}).  This can be accounted for by the addition of
a term $\int_{\bf r}\gamma (\hat {\bf e_h} \cdot \hat {\bf n})^2$ to the
Hamiltonian, where $\gamma$ is a phenomenological parameter, which we
assume to be a monotonically increasing function of $\rho_A$ and
strain. The uniaxial stretch also breaks both the hexagonal symmetry
of the lattice, and its rotation invariance.  Thus, we are forced to
consider both a more general lattice structure and an elastic
Hamiltonian that need not, and indeed, will not be invariant under
rotations of the lattice. We can take into account both of these
effects by replacing the harmonic pieces of the hexagonal elastic 
energy $\int_{\bf r}
({1\over2}\lambda u_{ii}^\perp u_{jj}^\perp + \mu u_{ij}^\perp
u_{ij}^\perp)$ with the more general harmonic elastic energy ${1 \over
  2}\int_{\bf r} C_{ijkl} \partial^{\perp}_i u_j \partial^{\perp}_k
u_l$, where the elastic constant tensor $C_{ijkl}$ is {\it not}
symmetric under interchange of its first two or second two indices,
due to the lack of in-plane rotation invariance just discussed.
The terms cubic and quartic in $\bf u$ in Eq.\ref{H} must also be so generalized of course.
Because of the $B_\perp$ term in Eq.\ref{H}, fluctuations of $\hat{\bf n}$ 
from the local column tangent $ \hat {\bf t} $ 
are small, i.e., $\delta{\bf n} \approx \partial_z \bf u$. 

We have analysed\cite{SJRT} this generalization of the model, Eq.\ref{H}, using
renormalization group (RG) methods\cite{RTpr}. One of the most surprising 
conclusions of this analysis is that, at long length scales, fluctuations 
$u_h \equiv \hat{\bf e}_h \cdot  {\bf u}$ along the direction of stretch decouple 
from those $u_s \equiv \hat{\bf e}_s \cdot  {\bf u}$ orthogonal to this direction, where we
have denoted the axis of stretch, $\hat {\bf e}_h$, as ``$\em hard$''($\em h$)
and called the other $\perp$ axis, orthogonal to $\hat {\bf e}_h$ ``$\em
soft$''($\em s$), i.e. ${\bf r}_\perp=( r_h, r_s)$. That is, all couplings
between $u_s$ and $u_h$ that are present in the full elastic tensor $C_{ijkl}$
flow to zero upon renormalization, leaving $C_{ijkl}$ in the form $C_{ijkl}
=B_{ss} \delta_{is} \delta_{js} \delta_{ks} \delta_{ls} + B_{sh} \delta_{is}
\delta_{jh} \delta_{ks} \delta_{lh} + B_{hs} \delta_{ih} \delta_{js}
\delta_{kh} \delta_{ls} + B_{hh} \delta_{ih} \delta_{jh} \delta_{kh}
\delta_{lh}$.

The total Hamiltonian for the system can therefore be expressed as a
sum of {\em decoupled} Hamiltonians for $u_h$ and $u_s$: $H_{tot}[u_h ,
u_s]=H_{XY}[u_h] + H_{m=1}[u_s]$, with
\begin{mathletters}
\begin{eqnarray}
H_{XY}&=&{1 \over 2}\int_{\bf r} \bigg[ \gamma|\partial_z u_h|^2 
+B_{sh}|\partial_s u_h|^2 + B_{hh}|\partial_h u_h|^2\nonumber\\ 
&+&{\cal R}e \sum_i V_i({\bf r})e^{i ({\bf G}_i 
\cdot \hat{\bf e}_h) u_h({\bf r})}\bigg]\;,
\label{H_XY}\\
H_{m=1} &=&  \int_{\bf r}\bigg[  {K\over2}
(\partial_z^2 u_s)^2 + B_{ss}\Big(\partial_s u_s - {1\over
2}(\partial_z u_s)^2\Big)^2 \nonumber\\ 
&+& B_{hs}(\partial_h u_s)^2 - 
g_z ({\bf g} \cdot \hat{\bf e}_s)(\partial_z u_s)\bigg]\;.
\label{H_m=1r}
\end{eqnarray}
\end{mathletters}
These two parts of $H_{tot}$, which describe the fluctuations of the
``hard'' and ``soft'' phonon fields $u_h$ and $u_s$, are not new.
Hamiltonians of precisely this form have been previously used to
describe the random field XY-model\cite{RFXY} and the ``m=1 smectic
Bragg glass''\cite{JSRT}, respectively, and have been studied
extensively.  However, a columnar phase confined in
anisotropic aerogel, whose Hamiltonian, $H_{tot}$, is a combination, or
{\em hybrid}, of the two, is entirely novel.

The $u_h$ fluctuations of our system are the same as those of a
slightly anisotropic (although isotropic in {\em scaling}) random
field XY-model and are given by $\overline{\langle(u_h({\bf
    r})-u_h({\bf 0} ))^2\rangle}= C(d)\ln{r}/G_{0h}^2$, where $G_{0h}$ is the
lattice spacing of the projection of the discotic reciprocal lattice
onto the hard axis.
%, and we've defined 
%$r' \equiv \sqrt{(r_s/a_s)^2+(r_h/a_h)^2+(r_z/a_z)^2}$.
They diverge logarithmically as a function of distance, implying that
the translational order along the hard direction is quasi-long-ranged.
While these elastic distortions are reminiscent of the famous
Landau-Peierls $\ln{r}$ fluctuations of {\em bulk} smectics they
differ crucially in two ways. Firstly, they are $\em disorder$, rather
than $\em thermally$, driven with $C(d)$ {\em universal} ($C(3)\approx 1.1$)
and the logarithm persisting in all $2<d<4$. Secondly, they are $\em
isotropic$ in their scaling. In contrast, in a bulk smectic the layer
fluctuations within the layers scale differently than those along the
normal to the layers.
%Here, the only anisotropy is in the different values of 
%$a_{s,h,z}$. 

In $H_{m=1}$ the combination of relevant anharmonic terms and large
disorder-induced $u_s$ fluctuations leads to strong anomalous
elasticity\cite{RTpr,SJRT}. By anomalous elasticity we mean that
the full, anharmonic theory with {\it constant} $K$,
$B_{ss}$ and $\Delta$ can, at small wavevector $k \ll \xi_{NL}^{-1}$ 
(where $\xi_{NL}$ is a non-universal length determined by material parameters, 
e.g., aerogel density), be effectively replaced by a harmonic theory with 
wavevector dependent $K$, $B_{ss}$ and $\Delta$. These are given by
\begin{mathletters}
\begin{eqnarray}
K({\bf k})&=&K k_z^{-\eta_K}
f_K\left({k_h/k_z^{\zeta_h}}, {k_s/k_z^{\zeta_s}}\right)\;,
\label{K}\\ 
B_{ss}({\bf k})&=&B_{ss} k_z^{\eta_B}
f_B\left({k_h/k_z^{\zeta_h}}, {k_s/k_z^{\zeta_s}}\right)\;,
\label{B_{ss}}\\ 
\Delta({\bf k})&=&\Delta k_z^{-\eta_{\Delta}}
f_{\Delta}\left({k_h/k_z^{\zeta_h}}, {k_s/k_z^{\zeta_s}}\right)\;.
\label{Delta}
\end{eqnarray}
\label{anom_elasticity}
\end{mathletters}
\noindent$B_{hs}$ is
not significantly renormalized; that is $B_{hs}({\bf k})=B_{hs} , $
independent of wavevector.  Here the anisotropy exponents
$\zeta_s\equiv 2-(\eta_B+\eta_K)/2$ and $\zeta_h\equiv 2-\eta_K/2$.
The exponents, evaluated using the renormalization group and a high
precision $\epsilon$-expansion were found to be: $\eta_K=0.50$,
$\eta_B=0.26$, and $\eta_{\Delta}=0.13$.\cite{SJRT,JSRT} We also
predict that the anomalous exponents will obey the following {\em exact} scaling relation:
\begin{eqnarray}
1+\eta_\Delta=\eta_B/2 + 2\eta_K .
\label{escale}
\end{eqnarray}
The translational order of the system along the soft direction is
short-ranged and is characterized by the algebraic and anisotropic
divergence of $u_s$ correlations,
%$\overline{\langle(\delta u_s({\bf r}))^2\rangle}$
%
\begin{mathletters}
\begin{eqnarray}
\overline{\langle(\delta u_s({\bf r}))^2\rangle}
&=&\cases{({K\over B_{ss}})({ r_z\over\xi_z})^{\chi_z}, &
$r_z \gg r_{s,h}$\cr
({K\over B_{ss}})({ r_s\over \xi_z^2}({K\over B_{ss}})^{1/2})^{{\chi_s
}}, & 
$r_s \gg r_{h,z}$\cr 
({K\over B_{ss}})({ r_h\over \xi_z^2}({K\over B_{sh}})^{1/2})^{{\chi_h}}, &
$r_h \gg r_{s,z}$}\nonumber\\
\label{xiX}
\end{eqnarray}
\end{mathletters}
where we have defined $\delta u_s({\bf r}) \equiv u_s({\bf r})-u_s(0)$, $\chi_z \equiv
1-\eta_K+\eta_B/2+\eta_{\Delta} = \eta_B + \eta_K $, $\chi_{s ,
h} \equiv {\chi_z/\zeta_{s , h}}$ , and $\xi_z\equiv {K^2
B_{sh}^{1/2}/(\Delta B_{ss}^{1/2})}$. In the second equality for
$\chi_z$, we have used the {\em exact} scaling relation between the
$\eta$'s. The exact scaling relation Eq.\ref{escale}, could be experimentally tested by
using the more general expressions for $\chi_z$, $\chi_s$, $\chi_h$ in terms of all three
exponents, and verifying that $\eta_B, \eta_K, and \eta_{\Delta}$ obey Eq.\ref{escale}.
Our $\epsilon$-expansion results for the $\eta$'s imply $\chi_z = 0.76$ , $\chi_s = 0.47$ 
and $\chi_h = 0.43$. The fluctuations given in
Eq.\ref{xiX}, like those along the hard direction, are disorder-,
rather than thermally-driven.

Despite this lack of translational order, our detailed
calculations\cite{SJRT} indicate that dislocation loops remain bound for
weak disorder, and therefore the low temperature phase replacing the
columnar phase must be distinct from the smectic and hexatic,
separated from them by a thermodynamically sharp dislocation unbinding
phase transition.

The stability of this exotic glass phase is contingent upon our
assumption of long-ranged orientational order.  We validate this
assumption by calculating $\overline{\langle| {\bf n}({\bf
    r})- {\bf n}({\bf 0})|^2\rangle} = \overline{\langle|\partial_z
  {\bf u}({\bf r})-\partial_z {\bf u}({\bf 0})|^2\rangle}$ and showing that
it does {\it not} diverge as ${\bf r} \rightarrow \infty$ \cite{SJRT}.  
Although equilibration into the ground state might be slow and therefore
require field alignment, this
orientational order would allow experimentalists to investigate {\em
 single domain} samples of HCBG. The anisotropic scaling information
which is usually lost in a powder averaged x-ray scattering experiment
{\em would be retained in a single domain experiment} 
allowing detailed tests of our predictions for 
$\eta_K$, $\eta_B$ and $\eta_\Delta$.

The scattering pattern in the $\perp$ plane, obtained from a single
domain sample, would consist of a set of spots rather than the set of
rings that one would expect from a powder sample. This
pattern depends crucially on the relative orientations
within the $\perp$ plane of the reciprocal lattice and the axis of
stretch, $\hat {\bf e}_h$, which could vary from discotic to discotic since it
depends on the microscopic interactions between the discs and strands. 
The intensity of a Bragg spot at a
reciprocal lattice vector ${\bf G}$
\begin{mathletters}
\begin{eqnarray}
I({\bf G})&\propto&\int_{\bf r} \exp[-\overline{\langle[{\bf G}\cdot 
({\bf u}({\bf r})-{\bf u}({\bf 0}))]^2\rangle}/2]\,
\label{I(G)a}\\
&\propto& \int_{\bf r}\exp\big\{-\big[G_h^2\overline{\langle(u_h({\bf r})-u_h({\bf 0}))^2\rangle} \nonumber\\
&+& G_s^2\overline{\langle(u_s({\bf r})-u_s({\bf 0}))^2\rangle}\big]/2\big\}\,
\label{I(G)b}
\end{eqnarray}
\end{mathletters}
Unless $G_s=0$, the algebraically diverging $u_s$ fluctuations
dominate the logarithmically diverging $u_h$ fluctuations and the
integrand is exponentially damped, leading to an anisotropically
broadened Bragg peak.  If however, $G_s=0$ then the exponential
becomes $r^{-0.55 n^2}$, where $n=G_h/G_{0h}$, with $G_{0h}$ being the
magnitude of the {\it smallest} ${\bf G}$ lying {\em on} the hard
axis, and for $n<3$ the integral diverges as $r \rightarrow \infty$,
leading to quasi-sharp peaks for those $n$'s.  We therefore predict
two classes of hybrid columnar Bragg
glasses. The first, which we call a {\em commensurate} HCBG, has some
reciprocal lattice vectors that lie {\em along} the hard axis, and
will exhibit a scattering pattern with peaks lying on the hard axis,
with the first two quasi-sharp.  In the second, {\em
  incommensurate} HCBG, class, all the peaks lie off the hard axis,
and are anisotropically broadened by the contribution from the $u_s$
fluctuations.  The smectic-like scattering pattern
(Fig.\ref{scattering}), with a quasi-sharp peak on the first ring,
will therefore only be observed for commensurate HCBG's.

The dependence of the anisotropically broadened peak widths on the
bare elastic constants is the same for both classes of HCBG.  Setting
the $u-u$ correlation functions, given in Eq.\ref{xiX}, equal to
$G_s^{-2}$ and solving for $(r_z=\xi_z^{X})^{ -1}$, $(r_s=\xi_s^{X})^{
  -1}$ and $(r_h=\xi_h^{X})^{ -1}$ gives the width of the peak at $\bf
G$ along each of the $z$, $s$ and $h$ directions:
\begin{mathletters}
\begin{eqnarray}
(\xi_z^{X})^{ -1}&=&\xi_z^{-1}({G_s}^2 K/B_{ss})^{\chi_z^{-1}}\;,
\label{z}\\ 
(\xi_s^{X})^{ -1}&=&\xi_z^{-2} (K/B_{ss})^{1/2}({G_s}^2 K/B_{ss})^{\chi_s^{-1}}
\;,
\label{s}\\ 
(\xi_h^{X})^{ -1}&=&\xi_z^{-2} (K/B_{sh})^{1/2}({G_s}^2 K/B_{ss})^{\chi_h^{-1}}\;.
\label{h}
\end{eqnarray}
\label{peakwidths}
\end{mathletters}
The temperature dependence of $\xi^X_{z,s,h}$ could be used to determine
the exponents $\eta_K$, $\eta_B$ and $\eta_{\Delta}$ since the {\em bulk}
$K(T)$, $B_{ss}(T)$ and $B_{sh}(T)$ in Eq.\ref{peakwidths} have temperature
dependences that can be extracted from data on bulk materials.  A more
direct way to observe the anomalous elasticity would be a direct measurement
of the $u-u$ correlation function
$I({\bf q})\propto\overline{\langle | u_s (\delta {\bf q} ) |^2\rangle}$, 
which can be obtained\cite{SJRT} for {\it 
large} ${\bf q}$ (i.e., {\bf q}'s with at least one component bigger than the
corresponding inverse x-ray correlation length quoted above) by looking at an
intermediate regime in the ``tails'' of the broad x-ray scattering peaks. In
those tails, (i.e., for ${\bf q}={\bf G}+ \delta {\bf q}$ with $\xi_{NL}^{-1}
\gg | \delta q_\alpha | \gg (\xi^\chi_\alpha)^{-1} $ for at least one
Cartesian direction $\alpha = (h,s,z)$ \cite{N}), it can be shown that
\begin{eqnarray}
I({\bf q})&\propto&  {\Delta ({\bf \delta q} ) q_z^2
\over (B_{ss} ({\bf \delta q} ) \delta q_s^2 + K({\bf \delta q} ) \delta q_z^4 + B_{hs} \delta q_h^2)^2 } \;.
\label{Itail}
\end{eqnarray}
Hence, the {\bf q} dependence of $B_{ss}({\bf q} )$, $K({\bf q} )$
and $\Delta ({\bf
  q} )$ in Eq.\ref{anom_elasticity} could be tested directly by a fit
of scattering data to these tails.

A related experimental approach, which has the advantage of {\it not}
being restricted to wavevectors larger than the inverse x-ray
correlation lengths, but can, rather, explore arbitrarily small {\bf
  q}'s, is light scattering, which measures director fluctuations.
These can be related to the $u-u$ correlations via our condition
$\delta{\bf n} \approx \partial_z \bf u$. This yields
\begin{mathletters}
\begin{eqnarray}
\overline{\langle| \delta n_s ( {\bf q} ) |^2\rangle} &=&
{\Delta ({\bf q} ) q_z^4 \over (B_{ss} ({\bf q} ) 
q_s^2 + K({\bf q} ) q_z^4 + B_{hs} q_h^2)^2 } \;,
\label{delns}\\
\overline{\langle| \delta n_h ( {\bf q} ) |^2\rangle}
&=&\cases{C(3)/2 G_{0h}^2 q_z^2/q^3, & commen.\cr k_B T q_z^2 G({\bf
q}) +\Delta_h(\bf q) q_z^4 G({\bf q})^2, & incommen.}\nonumber\\
\label{delnh}
\end{eqnarray}
\end{mathletters}
where $\Delta_h(\bf q)$ in Eq.\ref{delnh} is renormalized, ${\bf q}$
dependent tilt disorder variance obeying the scaling law
$\Delta_h({\bf q})=\Delta_h q_z^{-(\eta_{\Delta}+\eta_B)}
f_{\Delta_h}\left({q_h/q_z^{\zeta_h}}, {q_s/q_z^{\zeta_s}}\right)$ and
$G({\bf q})=1/(B_{sh} q_s^2 + \gamma q_z^2 + B_{hh} q_h^2)$.  The
commensurate and incommensurate cases differ because in the
commensurate case there {\it is} a random field acting on $u_h$, while
in the incommensurate case there is no random field, leaving the
random tilt as the dominant disorder.

L.R. acknowledges support by the NSF DMR-9625111, the MRSEC
DMR-9809555, and the Sloan and Packard Foundations. J.T. and K.S. were
supported by the NSF DMR-9634596 and DMR-9980123.

\end{multicols}

\begin{references}
\bibitem{N} We have shown in \cite{SJRT} that the ratio 
${\xi_{NL}}/{\xi^\chi_\alpha}\rightarrow 0$ as the aerogel density
goes to zero. Hence, one could make the width of this window in 
$\delta q_\alpha$ as large as one likes simply by making the aerogel 
density sufficiently small.
\end{references}
\end{document}